\documentclass[useAMS,usenatbib,twocolumn]{mn2e}
\usepackage{natbib,graphics,epsfig}
\usepackage{amsmath}
\usepackage{amssymb}
\usepackage{textcomp}
\usepackage{graphicx}
\usepackage{color}
  \definecolor{snow}{rgb}{1.000000,0.980392,0.980392}
  \definecolor{ghost white}{rgb}{0.972549,0.972549,1.000000}
  \definecolor{GhostWhite}{rgb}{0.972549,0.972549,1.000000}
  \definecolor{white smoke}{rgb}{0.960784,0.960784,0.960784}
  \definecolor{WhiteSmoke}{rgb}{0.960784,0.960784,0.960784}
  \definecolor{gainsboro}{rgb}{0.862745,0.862745,0.862745}
  \definecolor{floral white}{rgb}{1.000000,0.980392,0.941176}
  \definecolor{FloralWhite}{rgb}{1.000000,0.980392,0.941176}
  \definecolor{old lace}{rgb}{0.992157,0.960784,0.901961}
  \definecolor{OldLace}{rgb}{0.992157,0.960784,0.901961}
  \definecolor{linen}{rgb}{0.980392,0.941176,0.901961}
  \definecolor{antique white}{rgb}{0.980392,0.921569,0.843137}
  \definecolor{AntiqueWhite}{rgb}{0.980392,0.921569,0.843137}
  \definecolor{papaya whip}{rgb}{1.000000,0.937255,0.835294}
  \definecolor{PapayaWhip}{rgb}{1.000000,0.937255,0.835294}
  \definecolor{blanched almond}{rgb}{1.000000,0.921569,0.803922}
  \definecolor{BlanchedAlmond}{rgb}{1.000000,0.921569,0.803922}
  \definecolor{bisque}{rgb}{1.000000,0.894118,0.768627}
  \definecolor{peach puff}{rgb}{1.000000,0.854902,0.725490}
  \definecolor{PeachPuff}{rgb}{1.000000,0.854902,0.725490}
  \definecolor{navajo white}{rgb}{1.000000,0.870588,0.678431}
  \definecolor{NavajoWhite}{rgb}{1.000000,0.870588,0.678431}
  \definecolor{moccasin}{rgb}{1.000000,0.894118,0.709804}
  \definecolor{cornsilk}{rgb}{1.000000,0.972549,0.862745}
  \definecolor{ivory}{rgb}{1.000000,1.000000,0.941176}
  \definecolor{lemon chiffon}{rgb}{1.000000,0.980392,0.803922}
  \definecolor{LemonChiffon}{rgb}{1.000000,0.980392,0.803922}
  \definecolor{seashell}{rgb}{1.000000,0.960784,0.933333}
  \definecolor{honeydew}{rgb}{0.941176,1.000000,0.941176}
  \definecolor{mint cream}{rgb}{0.960784,1.000000,0.980392}
  \definecolor{MintCream}{rgb}{0.960784,1.000000,0.980392}
  \definecolor{azure}{rgb}{0.941176,1.000000,1.000000}
  \definecolor{alice blue}{rgb}{0.941176,0.972549,1.000000}
  \definecolor{AliceBlue}{rgb}{0.941176,0.972549,1.000000}
  \definecolor{lavender}{rgb}{0.901961,0.901961,0.980392}
  \definecolor{lavender blush}{rgb}{1.000000,0.941176,0.960784}
  \definecolor{LavenderBlush}{rgb}{1.000000,0.941176,0.960784}
  \definecolor{misty rose}{rgb}{1.000000,0.894118,0.882353}
  \definecolor{MistyRose}{rgb}{1.000000,0.894118,0.882353}
  \definecolor{white}{rgb}{1.000000,1.000000,1.000000}
  \definecolor{black}{rgb}{0.000000,0.000000,0.000000}
  \definecolor{dark slate gray}{rgb}{0.184314,0.309804,0.309804}
  \definecolor{DarkSlateGray}{rgb}{0.184314,0.309804,0.309804}
  \definecolor{dark slate grey}{rgb}{0.184314,0.309804,0.309804}
  \definecolor{DarkSlateGrey}{rgb}{0.184314,0.309804,0.309804}
  \definecolor{dim gray}{rgb}{0.411765,0.411765,0.411765}
  \definecolor{DimGray}{rgb}{0.411765,0.411765,0.411765}
  \definecolor{dim grey}{rgb}{0.411765,0.411765,0.411765}
  \definecolor{DimGrey}{rgb}{0.411765,0.411765,0.411765}
  \definecolor{slate gray}{rgb}{0.439216,0.501961,0.564706}
  \definecolor{SlateGray}{rgb}{0.439216,0.501961,0.564706}
  \definecolor{slate grey}{rgb}{0.439216,0.501961,0.564706}
  \definecolor{SlateGrey}{rgb}{0.439216,0.501961,0.564706}
  \definecolor{light slate gray}{rgb}{0.466667,0.533333,0.600000}
  \definecolor{LightSlateGray}{rgb}{0.466667,0.533333,0.600000}
  \definecolor{light slate grey}{rgb}{0.466667,0.533333,0.600000}
  \definecolor{LightSlateGrey}{rgb}{0.466667,0.533333,0.600000}
  \definecolor{gray}{rgb}{0.745098,0.745098,0.745098}
  \definecolor{grey}{rgb}{0.745098,0.745098,0.745098}
  \definecolor{light grey}{rgb}{0.827451,0.827451,0.827451}
  \definecolor{LightGrey}{rgb}{0.827451,0.827451,0.827451}
  \definecolor{light gray}{rgb}{0.827451,0.827451,0.827451}
  \definecolor{LightGray}{rgb}{0.827451,0.827451,0.827451}
  \definecolor{midnight blue}{rgb}{0.098039,0.098039,0.439216}
  \definecolor{MidnightBlue}{rgb}{0.098039,0.098039,0.439216}
  \definecolor{navy}{rgb}{0.000000,0.000000,0.501961}
  \definecolor{navy blue}{rgb}{0.000000,0.000000,0.501961}
  \definecolor{NavyBlue}{rgb}{0.000000,0.000000,0.501961}
  \definecolor{cornflower blue}{rgb}{0.392157,0.584314,0.929412}
  \definecolor{CornflowerBlue}{rgb}{0.392157,0.584314,0.929412}
  \definecolor{dark slate blue}{rgb}{0.282353,0.239216,0.545098}
  \definecolor{DarkSlateBlue}{rgb}{0.282353,0.239216,0.545098}
  \definecolor{slate blue}{rgb}{0.415686,0.352941,0.803922}
  \definecolor{SlateBlue}{rgb}{0.415686,0.352941,0.803922}
  \definecolor{medium slate blue}{rgb}{0.482353,0.407843,0.933333}
  \definecolor{MediumSlateBlue}{rgb}{0.482353,0.407843,0.933333}
  \definecolor{light slate blue}{rgb}{0.517647,0.439216,1.000000}
  \definecolor{LightSlateBlue}{rgb}{0.517647,0.439216,1.000000}
  \definecolor{medium blue}{rgb}{0.000000,0.000000,0.803922}
  \definecolor{MediumBlue}{rgb}{0.000000,0.000000,0.803922}
  \definecolor{royal blue}{rgb}{0.254902,0.411765,0.882353}
  \definecolor{RoyalBlue}{rgb}{0.254902,0.411765,0.882353}
  \definecolor{blue}{rgb}{0.000000,0.000000,1.000000}
  \definecolor{dodger blue}{rgb}{0.117647,0.564706,1.000000}
  \definecolor{DodgerBlue}{rgb}{0.117647,0.564706,1.000000}
  \definecolor{deep sky blue}{rgb}{0.000000,0.749020,1.000000}
  \definecolor{DeepSkyBlue}{rgb}{0.000000,0.749020,1.000000}
  \definecolor{sky blue}{rgb}{0.529412,0.807843,0.921569}
  \definecolor{SkyBlue}{rgb}{0.529412,0.807843,0.921569}
  \definecolor{light sky blue}{rgb}{0.529412,0.807843,0.980392}
  \definecolor{LightSkyBlue}{rgb}{0.529412,0.807843,0.980392}
  \definecolor{steel blue}{rgb}{0.274510,0.509804,0.705882}
  \definecolor{SteelBlue}{rgb}{0.274510,0.509804,0.705882}
  \definecolor{light steel blue}{rgb}{0.690196,0.768627,0.870588}
  \definecolor{LightSteelBlue}{rgb}{0.690196,0.768627,0.870588}
  \definecolor{light blue}{rgb}{0.678431,0.847059,0.901961}
  \definecolor{LightBlue}{rgb}{0.678431,0.847059,0.901961}
  \definecolor{powder blue}{rgb}{0.690196,0.878431,0.901961}
  \definecolor{PowderBlue}{rgb}{0.690196,0.878431,0.901961}
  \definecolor{pale turquoise}{rgb}{0.686275,0.933333,0.933333}
  \definecolor{PaleTurquoise}{rgb}{0.686275,0.933333,0.933333}
  \definecolor{dark turquoise}{rgb}{0.000000,0.807843,0.819608}
  \definecolor{DarkTurquoise}{rgb}{0.000000,0.807843,0.819608}
  \definecolor{medium turquoise}{rgb}{0.282353,0.819608,0.800000}
  \definecolor{MediumTurquoise}{rgb}{0.282353,0.819608,0.800000}
  \definecolor{turquoise}{rgb}{0.250980,0.878431,0.815686}
  \definecolor{cyan}{rgb}{0.000000,1.000000,1.000000}
  \definecolor{light cyan}{rgb}{0.878431,1.000000,1.000000}
  \definecolor{LightCyan}{rgb}{0.878431,1.000000,1.000000}
  \definecolor{cadet blue}{rgb}{0.372549,0.619608,0.627451}
  \definecolor{CadetBlue}{rgb}{0.372549,0.619608,0.627451}
  \definecolor{medium aquamarine}{rgb}{0.400000,0.803922,0.666667}
  \definecolor{MediumAquamarine}{rgb}{0.400000,0.803922,0.666667}
  \definecolor{aquamarine}{rgb}{0.498039,1.000000,0.831373}
  \definecolor{dark green}{rgb}{0.000000,0.392157,0.000000}
  \definecolor{DarkGreen}{rgb}{0.000000,0.392157,0.000000}
  \definecolor{dark olive green}{rgb}{0.333333,0.419608,0.184314}
  \definecolor{DarkOliveGreen}{rgb}{0.333333,0.419608,0.184314}
  \definecolor{dark sea green}{rgb}{0.560784,0.737255,0.560784}
  \definecolor{DarkSeaGreen}{rgb}{0.560784,0.737255,0.560784}
  \definecolor{sea green}{rgb}{0.180392,0.545098,0.341176}
  \definecolor{SeaGreen}{rgb}{0.180392,0.545098,0.341176}
  \definecolor{medium sea green}{rgb}{0.235294,0.701961,0.443137}
  \definecolor{MediumSeaGreen}{rgb}{0.235294,0.701961,0.443137}
  \definecolor{light sea green}{rgb}{0.125490,0.698039,0.666667}
  \definecolor{LightSeaGreen}{rgb}{0.125490,0.698039,0.666667}
  \definecolor{pale green}{rgb}{0.596078,0.984314,0.596078}
  \definecolor{PaleGreen}{rgb}{0.596078,0.984314,0.596078}
  \definecolor{spring green}{rgb}{0.000000,1.000000,0.498039}
  \definecolor{SpringGreen}{rgb}{0.000000,1.000000,0.498039}
  \definecolor{lawn green}{rgb}{0.486275,0.988235,0.000000}
  \definecolor{LawnGreen}{rgb}{0.486275,0.988235,0.000000}
  \definecolor{green}{rgb}{0.000000,1.000000,0.000000}
  \definecolor{chartreuse}{rgb}{0.498039,1.000000,0.000000}
  \definecolor{medium spring green}{rgb}{0.000000,0.980392,0.603922}
  \definecolor{MediumSpringGreen}{rgb}{0.000000,0.980392,0.603922}
  \definecolor{green yellow}{rgb}{0.678431,1.000000,0.184314}
  \definecolor{GreenYellow}{rgb}{0.678431,1.000000,0.184314}
  \definecolor{lime green}{rgb}{0.196078,0.803922,0.196078}
  \definecolor{LimeGreen}{rgb}{0.196078,0.803922,0.196078}
  \definecolor{yellow green}{rgb}{0.603922,0.803922,0.196078}
  \definecolor{YellowGreen}{rgb}{0.603922,0.803922,0.196078}
  \definecolor{forest green}{rgb}{0.133333,0.545098,0.133333}
  \definecolor{ForestGreen}{rgb}{0.133333,0.545098,0.133333}
  \definecolor{olive drab}{rgb}{0.419608,0.556863,0.137255}
  \definecolor{OliveDrab}{rgb}{0.419608,0.556863,0.137255}
  \definecolor{dark khaki}{rgb}{0.741176,0.717647,0.419608}
  \definecolor{DarkKhaki}{rgb}{0.741176,0.717647,0.419608}
  \definecolor{khaki}{rgb}{0.941176,0.901961,0.549020}
  \definecolor{pale goldenrod}{rgb}{0.933333,0.909804,0.666667}
  \definecolor{PaleGoldenrod}{rgb}{0.933333,0.909804,0.666667}
  \definecolor{light goldenrod yellow}{rgb}{0.980392,0.980392,0.823529}
  \definecolor{LightGoldenrodYellow}{rgb}{0.980392,0.980392,0.823529}
  \definecolor{light yellow}{rgb}{1.000000,1.000000,0.878431}
  \definecolor{LightYellow}{rgb}{1.000000,1.000000,0.878431}
  \definecolor{yellow}{rgb}{1.000000,1.000000,0.000000}
  \definecolor{gold}{rgb}{1.000000,0.843137,0.000000}
  \definecolor{light goldenrod}{rgb}{0.933333,0.866667,0.509804}
  \definecolor{LightGoldenrod}{rgb}{0.933333,0.866667,0.509804}
  \definecolor{goldenrod}{rgb}{0.854902,0.647059,0.125490}
  \definecolor{dark goldenrod}{rgb}{0.721569,0.525490,0.043137}
  \definecolor{DarkGoldenrod}{rgb}{0.721569,0.525490,0.043137}
  \definecolor{rosy brown}{rgb}{0.737255,0.560784,0.560784}
  \definecolor{RosyBrown}{rgb}{0.737255,0.560784,0.560784}
  \definecolor{indian red}{rgb}{0.803922,0.360784,0.360784}
  \definecolor{IndianRed}{rgb}{0.803922,0.360784,0.360784}
  \definecolor{saddle brown}{rgb}{0.545098,0.270588,0.074510}
  \definecolor{SaddleBrown}{rgb}{0.545098,0.270588,0.074510}
  \definecolor{sienna}{rgb}{0.627451,0.321569,0.176471}
  \definecolor{peru}{rgb}{0.803922,0.521569,0.247059}
  \definecolor{burlywood}{rgb}{0.870588,0.721569,0.529412}
  \definecolor{beige}{rgb}{0.960784,0.960784,0.862745}
  \definecolor{wheat}{rgb}{0.960784,0.870588,0.701961}
  \definecolor{sandy brown}{rgb}{0.956863,0.643137,0.376471}
  \definecolor{SandyBrown}{rgb}{0.956863,0.643137,0.376471}
  \definecolor{tan}{rgb}{0.823529,0.705882,0.549020}
  \definecolor{chocolate}{rgb}{0.823529,0.411765,0.117647}
  \definecolor{firebrick}{rgb}{0.698039,0.133333,0.133333}
  \definecolor{brown}{rgb}{0.647059,0.164706,0.164706}
  \definecolor{dark salmon}{rgb}{0.913725,0.588235,0.478431}
  \definecolor{DarkSalmon}{rgb}{0.913725,0.588235,0.478431}
  \definecolor{salmon}{rgb}{0.980392,0.501961,0.447059}
  \definecolor{light salmon}{rgb}{1.000000,0.627451,0.478431}
  \definecolor{LightSalmon}{rgb}{1.000000,0.627451,0.478431}
  \definecolor{orange}{rgb}{1.000000,0.647059,0.000000}
  \definecolor{dark orange}{rgb}{1.000000,0.549020,0.000000}
  \definecolor{DarkOrange}{rgb}{1.000000,0.549020,0.000000}
  \definecolor{coral}{rgb}{1.000000,0.498039,0.313726}
  \definecolor{light coral}{rgb}{0.941176,0.501961,0.501961}
  \definecolor{LightCoral}{rgb}{0.941176,0.501961,0.501961}
  \definecolor{tomato}{rgb}{1.000000,0.388235,0.278431}
  \definecolor{orange red}{rgb}{1.000000,0.270588,0.000000}
  \definecolor{OrangeRed}{rgb}{1.000000,0.270588,0.000000}
  \definecolor{red}{rgb}{1.000000,0.000000,0.000000}
  \definecolor{hot pink}{rgb}{1.000000,0.411765,0.705882}
  \definecolor{HotPink}{rgb}{1.000000,0.411765,0.705882}
  \definecolor{deep pink}{rgb}{1.000000,0.078431,0.576471}
  \definecolor{DeepPink}{rgb}{1.000000,0.078431,0.576471}
  \definecolor{pink}{rgb}{1.000000,0.752941,0.796078}
  \definecolor{light pink}{rgb}{1.000000,0.713726,0.756863}
  \definecolor{LightPink}{rgb}{1.000000,0.713726,0.756863}
  \definecolor{pale violet red}{rgb}{0.858824,0.439216,0.576471}
  \definecolor{PaleVioletRed}{rgb}{0.858824,0.439216,0.576471}
  \definecolor{maroon}{rgb}{0.690196,0.188235,0.376471}
  \definecolor{medium violet red}{rgb}{0.780392,0.082353,0.521569}
  \definecolor{MediumVioletRed}{rgb}{0.780392,0.082353,0.521569}
  \definecolor{violet red}{rgb}{0.815686,0.125490,0.564706}
  \definecolor{VioletRed}{rgb}{0.815686,0.125490,0.564706}
  \definecolor{magenta}{rgb}{1.000000,0.000000,1.000000}
  \definecolor{violet}{rgb}{0.933333,0.509804,0.933333}
  \definecolor{plum}{rgb}{0.866667,0.627451,0.866667}
  \definecolor{orchid}{rgb}{0.854902,0.439216,0.839216}
  \definecolor{medium orchid}{rgb}{0.729412,0.333333,0.827451}
  \definecolor{MediumOrchid}{rgb}{0.729412,0.333333,0.827451}
  \definecolor{dark orchid}{rgb}{0.600000,0.196078,0.800000}
  \definecolor{DarkOrchid}{rgb}{0.600000,0.196078,0.800000}
  \definecolor{dark violet}{rgb}{0.580392,0.000000,0.827451}
  \definecolor{DarkViolet}{rgb}{0.580392,0.000000,0.827451}
  \definecolor{blue violet}{rgb}{0.541176,0.168627,0.886275}
  \definecolor{BlueViolet}{rgb}{0.541176,0.168627,0.886275}
  \definecolor{purple}{rgb}{0.627451,0.125490,0.941176}
  \definecolor{medium purple}{rgb}{0.576471,0.439216,0.858824}
  \definecolor{MediumPurple}{rgb}{0.576471,0.439216,0.858824}
  \definecolor{thistle}{rgb}{0.847059,0.749020,0.847059}
  \definecolor{snow1}{rgb}{1.000000,0.980392,0.980392}
  \definecolor{snow2}{rgb}{0.933333,0.913725,0.913725}
  \definecolor{snow3}{rgb}{0.803922,0.788235,0.788235}
  \definecolor{snow4}{rgb}{0.545098,0.537255,0.537255}
  \definecolor{seashell1}{rgb}{1.000000,0.960784,0.933333}
  \definecolor{seashell2}{rgb}{0.933333,0.898039,0.870588}
  \definecolor{seashell3}{rgb}{0.803922,0.772549,0.749020}
  \definecolor{seashell4}{rgb}{0.545098,0.525490,0.509804}
  \definecolor{AntiqueWhite1}{rgb}{1.000000,0.937255,0.858824}
  \definecolor{AntiqueWhite2}{rgb}{0.933333,0.874510,0.800000}
  \definecolor{AntiqueWhite3}{rgb}{0.803922,0.752941,0.690196}
  \definecolor{AntiqueWhite4}{rgb}{0.545098,0.513726,0.470588}
  \definecolor{bisque1}{rgb}{1.000000,0.894118,0.768627}
  \definecolor{bisque2}{rgb}{0.933333,0.835294,0.717647}
  \definecolor{bisque3}{rgb}{0.803922,0.717647,0.619608}
  \definecolor{bisque4}{rgb}{0.545098,0.490196,0.419608}
  \definecolor{PeachPuff1}{rgb}{1.000000,0.854902,0.725490}
  \definecolor{PeachPuff2}{rgb}{0.933333,0.796078,0.678431}
  \definecolor{PeachPuff3}{rgb}{0.803922,0.686275,0.584314}
  \definecolor{PeachPuff4}{rgb}{0.545098,0.466667,0.396078}
  \definecolor{NavajoWhite1}{rgb}{1.000000,0.870588,0.678431}
  \definecolor{NavajoWhite2}{rgb}{0.933333,0.811765,0.631373}
  \definecolor{NavajoWhite3}{rgb}{0.803922,0.701961,0.545098}
  \definecolor{NavajoWhite4}{rgb}{0.545098,0.474510,0.368627}
  \definecolor{LemonChiffon1}{rgb}{1.000000,0.980392,0.803922}
  \definecolor{LemonChiffon2}{rgb}{0.933333,0.913725,0.749020}
  \definecolor{LemonChiffon3}{rgb}{0.803922,0.788235,0.647059}
  \definecolor{LemonChiffon4}{rgb}{0.545098,0.537255,0.439216}
  \definecolor{cornsilk1}{rgb}{1.000000,0.972549,0.862745}
  \definecolor{cornsilk2}{rgb}{0.933333,0.909804,0.803922}
  \definecolor{cornsilk3}{rgb}{0.803922,0.784314,0.694118}
  \definecolor{cornsilk4}{rgb}{0.545098,0.533333,0.470588}
  \definecolor{ivory1}{rgb}{1.000000,1.000000,0.941176}
  \definecolor{ivory2}{rgb}{0.933333,0.933333,0.878431}
  \definecolor{ivory3}{rgb}{0.803922,0.803922,0.756863}
  \definecolor{ivory4}{rgb}{0.545098,0.545098,0.513726}
  \definecolor{honeydew1}{rgb}{0.941176,1.000000,0.941176}
  \definecolor{honeydew2}{rgb}{0.878431,0.933333,0.878431}
  \definecolor{honeydew3}{rgb}{0.756863,0.803922,0.756863}
  \definecolor{honeydew4}{rgb}{0.513726,0.545098,0.513726}
  \definecolor{LavenderBlush1}{rgb}{1.000000,0.941176,0.960784}
  \definecolor{LavenderBlush2}{rgb}{0.933333,0.878431,0.898039}
  \definecolor{LavenderBlush3}{rgb}{0.803922,0.756863,0.772549}
  \definecolor{LavenderBlush4}{rgb}{0.545098,0.513726,0.525490}
  \definecolor{MistyRose1}{rgb}{1.000000,0.894118,0.882353}
  \definecolor{MistyRose2}{rgb}{0.933333,0.835294,0.823529}
  \definecolor{MistyRose3}{rgb}{0.803922,0.717647,0.709804}
  \definecolor{MistyRose4}{rgb}{0.545098,0.490196,0.482353}
  \definecolor{azure1}{rgb}{0.941176,1.000000,1.000000}
  \definecolor{azure2}{rgb}{0.878431,0.933333,0.933333}
  \definecolor{azure3}{rgb}{0.756863,0.803922,0.803922}
  \definecolor{azure4}{rgb}{0.513726,0.545098,0.545098}
  \definecolor{SlateBlue1}{rgb}{0.513726,0.435294,1.000000}
  \definecolor{SlateBlue2}{rgb}{0.478431,0.403922,0.933333}
  \definecolor{SlateBlue3}{rgb}{0.411765,0.349020,0.803922}
  \definecolor{SlateBlue4}{rgb}{0.278431,0.235294,0.545098}
  \definecolor{RoyalBlue1}{rgb}{0.282353,0.462745,1.000000}
  \definecolor{RoyalBlue2}{rgb}{0.262745,0.431373,0.933333}
  \definecolor{RoyalBlue3}{rgb}{0.227451,0.372549,0.803922}
  \definecolor{RoyalBlue4}{rgb}{0.152941,0.250980,0.545098}
  \definecolor{blue1}{rgb}{0.000000,0.000000,1.000000}
  \definecolor{blue2}{rgb}{0.000000,0.000000,0.933333}
  \definecolor{blue3}{rgb}{0.000000,0.000000,0.803922}
  \definecolor{blue4}{rgb}{0.000000,0.000000,0.545098}
  \definecolor{DodgerBlue1}{rgb}{0.117647,0.564706,1.000000}
  \definecolor{DodgerBlue2}{rgb}{0.109804,0.525490,0.933333}
  \definecolor{DodgerBlue3}{rgb}{0.094118,0.454902,0.803922}
  \definecolor{DodgerBlue4}{rgb}{0.062745,0.305882,0.545098}
  \definecolor{SteelBlue1}{rgb}{0.388235,0.721569,1.000000}
  \definecolor{SteelBlue2}{rgb}{0.360784,0.674510,0.933333}
  \definecolor{SteelBlue3}{rgb}{0.309804,0.580392,0.803922}
  \definecolor{SteelBlue4}{rgb}{0.211765,0.392157,0.545098}
  \definecolor{DeepSkyBlue1}{rgb}{0.000000,0.749020,1.000000}
  \definecolor{DeepSkyBlue2}{rgb}{0.000000,0.698039,0.933333}
  \definecolor{DeepSkyBlue3}{rgb}{0.000000,0.603922,0.803922}
  \definecolor{DeepSkyBlue4}{rgb}{0.000000,0.407843,0.545098}
  \definecolor{SkyBlue1}{rgb}{0.529412,0.807843,1.000000}
  \definecolor{SkyBlue2}{rgb}{0.494118,0.752941,0.933333}
  \definecolor{SkyBlue3}{rgb}{0.423529,0.650980,0.803922}
  \definecolor{SkyBlue4}{rgb}{0.290196,0.439216,0.545098}
  \definecolor{LightSkyBlue1}{rgb}{0.690196,0.886275,1.000000}
  \definecolor{LightSkyBlue2}{rgb}{0.643137,0.827451,0.933333}
  \definecolor{LightSkyBlue3}{rgb}{0.552941,0.713726,0.803922}
  \definecolor{LightSkyBlue4}{rgb}{0.376471,0.482353,0.545098}
  \definecolor{SlateGray1}{rgb}{0.776471,0.886275,1.000000}
  \definecolor{SlateGray2}{rgb}{0.725490,0.827451,0.933333}
  \definecolor{SlateGray3}{rgb}{0.623529,0.713726,0.803922}
  \definecolor{SlateGray4}{rgb}{0.423529,0.482353,0.545098}
  \definecolor{LightSteelBlue1}{rgb}{0.792157,0.882353,1.000000}
  \definecolor{LightSteelBlue2}{rgb}{0.737255,0.823529,0.933333}
  \definecolor{LightSteelBlue3}{rgb}{0.635294,0.709804,0.803922}
  \definecolor{LightSteelBlue4}{rgb}{0.431373,0.482353,0.545098}
  \definecolor{LightBlue1}{rgb}{0.749020,0.937255,1.000000}
  \definecolor{LightBlue2}{rgb}{0.698039,0.874510,0.933333}
  \definecolor{LightBlue3}{rgb}{0.603922,0.752941,0.803922}
  \definecolor{LightBlue4}{rgb}{0.407843,0.513726,0.545098}
  \definecolor{LightCyan1}{rgb}{0.878431,1.000000,1.000000}
  \definecolor{LightCyan2}{rgb}{0.819608,0.933333,0.933333}
  \definecolor{LightCyan3}{rgb}{0.705882,0.803922,0.803922}
  \definecolor{LightCyan4}{rgb}{0.478431,0.545098,0.545098}
  \definecolor{PaleTurquoise1}{rgb}{0.733333,1.000000,1.000000}
  \definecolor{PaleTurquoise2}{rgb}{0.682353,0.933333,0.933333}
  \definecolor{PaleTurquoise3}{rgb}{0.588235,0.803922,0.803922}
  \definecolor{PaleTurquoise4}{rgb}{0.400000,0.545098,0.545098}
  \definecolor{CadetBlue1}{rgb}{0.596078,0.960784,1.000000}
  \definecolor{CadetBlue2}{rgb}{0.556863,0.898039,0.933333}
  \definecolor{CadetBlue3}{rgb}{0.478431,0.772549,0.803922}
  \definecolor{CadetBlue4}{rgb}{0.325490,0.525490,0.545098}
  \definecolor{turquoise1}{rgb}{0.000000,0.960784,1.000000}
  \definecolor{turquoise2}{rgb}{0.000000,0.898039,0.933333}
  \definecolor{turquoise3}{rgb}{0.000000,0.772549,0.803922}
  \definecolor{turquoise4}{rgb}{0.000000,0.525490,0.545098}
  \definecolor{cyan1}{rgb}{0.000000,1.000000,1.000000}
  \definecolor{cyan2}{rgb}{0.000000,0.933333,0.933333}
  \definecolor{cyan3}{rgb}{0.000000,0.803922,0.803922}
  \definecolor{cyan4}{rgb}{0.000000,0.545098,0.545098}
  \definecolor{DarkSlateGray1}{rgb}{0.592157,1.000000,1.000000}
  \definecolor{DarkSlateGray2}{rgb}{0.552941,0.933333,0.933333}
  \definecolor{DarkSlateGray3}{rgb}{0.474510,0.803922,0.803922}
  \definecolor{DarkSlateGray4}{rgb}{0.321569,0.545098,0.545098}
  \definecolor{aquamarine1}{rgb}{0.498039,1.000000,0.831373}
  \definecolor{aquamarine2}{rgb}{0.462745,0.933333,0.776471}
  \definecolor{aquamarine3}{rgb}{0.400000,0.803922,0.666667}
  \definecolor{aquamarine4}{rgb}{0.270588,0.545098,0.454902}
  \definecolor{DarkSeaGreen1}{rgb}{0.756863,1.000000,0.756863}
  \definecolor{DarkSeaGreen2}{rgb}{0.705882,0.933333,0.705882}
  \definecolor{DarkSeaGreen3}{rgb}{0.607843,0.803922,0.607843}
  \definecolor{DarkSeaGreen4}{rgb}{0.411765,0.545098,0.411765}
  \definecolor{SeaGreen1}{rgb}{0.329412,1.000000,0.623529}
  \definecolor{SeaGreen2}{rgb}{0.305882,0.933333,0.580392}
  \definecolor{SeaGreen3}{rgb}{0.262745,0.803922,0.501961}
  \definecolor{SeaGreen4}{rgb}{0.180392,0.545098,0.341176}
  \definecolor{PaleGreen1}{rgb}{0.603922,1.000000,0.603922}
  \definecolor{PaleGreen2}{rgb}{0.564706,0.933333,0.564706}
  \definecolor{PaleGreen3}{rgb}{0.486275,0.803922,0.486275}
  \definecolor{PaleGreen4}{rgb}{0.329412,0.545098,0.329412}
  \definecolor{SpringGreen1}{rgb}{0.000000,1.000000,0.498039}
  \definecolor{SpringGreen2}{rgb}{0.000000,0.933333,0.462745}
  \definecolor{SpringGreen3}{rgb}{0.000000,0.803922,0.400000}
  \definecolor{SpringGreen4}{rgb}{0.000000,0.545098,0.270588}
  \definecolor{green1}{rgb}{0.000000,1.000000,0.000000}
  \definecolor{green2}{rgb}{0.000000,0.933333,0.000000}
  \definecolor{green3}{rgb}{0.000000,0.803922,0.000000}
  \definecolor{green4}{rgb}{0.000000,0.545098,0.000000}
  \definecolor{chartreuse1}{rgb}{0.498039,1.000000,0.000000}
  \definecolor{chartreuse2}{rgb}{0.462745,0.933333,0.000000}
  \definecolor{chartreuse3}{rgb}{0.400000,0.803922,0.000000}
  \definecolor{chartreuse4}{rgb}{0.270588,0.545098,0.000000}
  \definecolor{OliveDrab1}{rgb}{0.752941,1.000000,0.243137}
  \definecolor{OliveDrab2}{rgb}{0.701961,0.933333,0.227451}
  \definecolor{OliveDrab3}{rgb}{0.603922,0.803922,0.196078}
  \definecolor{OliveDrab4}{rgb}{0.411765,0.545098,0.133333}
  \definecolor{DarkOliveGreen1}{rgb}{0.792157,1.000000,0.439216}
  \definecolor{DarkOliveGreen2}{rgb}{0.737255,0.933333,0.407843}
  \definecolor{DarkOliveGreen3}{rgb}{0.635294,0.803922,0.352941}
  \definecolor{DarkOliveGreen4}{rgb}{0.431373,0.545098,0.239216}
  \definecolor{khaki1}{rgb}{1.000000,0.964706,0.560784}
  \definecolor{khaki2}{rgb}{0.933333,0.901961,0.521569}
  \definecolor{khaki3}{rgb}{0.803922,0.776471,0.450980}
  \definecolor{khaki4}{rgb}{0.545098,0.525490,0.305882}
  \definecolor{LightGoldenrod1}{rgb}{1.000000,0.925490,0.545098}
  \definecolor{LightGoldenrod2}{rgb}{0.933333,0.862745,0.509804}
  \definecolor{LightGoldenrod3}{rgb}{0.803922,0.745098,0.439216}
  \definecolor{LightGoldenrod4}{rgb}{0.545098,0.505882,0.298039}
  \definecolor{LightYellow1}{rgb}{1.000000,1.000000,0.878431}
  \definecolor{LightYellow2}{rgb}{0.933333,0.933333,0.819608}
  \definecolor{LightYellow3}{rgb}{0.803922,0.803922,0.705882}
  \definecolor{LightYellow4}{rgb}{0.545098,0.545098,0.478431}
  \definecolor{yellow1}{rgb}{1.000000,1.000000,0.000000}
  \definecolor{yellow2}{rgb}{0.933333,0.933333,0.000000}
  \definecolor{yellow3}{rgb}{0.803922,0.803922,0.000000}
  \definecolor{yellow4}{rgb}{0.545098,0.545098,0.000000}
  \definecolor{gold1}{rgb}{1.000000,0.843137,0.000000}
  \definecolor{gold2}{rgb}{0.933333,0.788235,0.000000}
  \definecolor{gold3}{rgb}{0.803922,0.678431,0.000000}
  \definecolor{gold4}{rgb}{0.545098,0.458824,0.000000}
  \definecolor{goldenrod1}{rgb}{1.000000,0.756863,0.145098}
  \definecolor{goldenrod2}{rgb}{0.933333,0.705882,0.133333}
  \definecolor{goldenrod3}{rgb}{0.803922,0.607843,0.113725}
  \definecolor{goldenrod4}{rgb}{0.545098,0.411765,0.078431}
  \definecolor{DarkGoldenrod1}{rgb}{1.000000,0.725490,0.058824}
  \definecolor{DarkGoldenrod2}{rgb}{0.933333,0.678431,0.054902}
  \definecolor{DarkGoldenrod3}{rgb}{0.803922,0.584314,0.047059}
  \definecolor{DarkGoldenrod4}{rgb}{0.545098,0.396078,0.031373}
  \definecolor{RosyBrown1}{rgb}{1.000000,0.756863,0.756863}
  \definecolor{RosyBrown2}{rgb}{0.933333,0.705882,0.705882}
  \definecolor{RosyBrown3}{rgb}{0.803922,0.607843,0.607843}
  \definecolor{RosyBrown4}{rgb}{0.545098,0.411765,0.411765}
  \definecolor{IndianRed1}{rgb}{1.000000,0.415686,0.415686}
  \definecolor{IndianRed2}{rgb}{0.933333,0.388235,0.388235}
  \definecolor{IndianRed3}{rgb}{0.803922,0.333333,0.333333}
  \definecolor{IndianRed4}{rgb}{0.545098,0.227451,0.227451}
  \definecolor{sienna1}{rgb}{1.000000,0.509804,0.278431}
  \definecolor{sienna2}{rgb}{0.933333,0.474510,0.258824}
  \definecolor{sienna3}{rgb}{0.803922,0.407843,0.223529}
  \definecolor{sienna4}{rgb}{0.545098,0.278431,0.149020}
  \definecolor{burlywood1}{rgb}{1.000000,0.827451,0.607843}
  \definecolor{burlywood2}{rgb}{0.933333,0.772549,0.568627}
  \definecolor{burlywood3}{rgb}{0.803922,0.666667,0.490196}
  \definecolor{burlywood4}{rgb}{0.545098,0.450980,0.333333}
  \definecolor{wheat1}{rgb}{1.000000,0.905882,0.729412}
  \definecolor{wheat2}{rgb}{0.933333,0.847059,0.682353}
  \definecolor{wheat3}{rgb}{0.803922,0.729412,0.588235}
  \definecolor{wheat4}{rgb}{0.545098,0.494118,0.400000}
  \definecolor{tan1}{rgb}{1.000000,0.647059,0.309804}
  \definecolor{tan2}{rgb}{0.933333,0.603922,0.286275}
  \definecolor{tan3}{rgb}{0.803922,0.521569,0.247059}
  \definecolor{tan4}{rgb}{0.545098,0.352941,0.168627}
  \definecolor{chocolate1}{rgb}{1.000000,0.498039,0.141176}
  \definecolor{chocolate2}{rgb}{0.933333,0.462745,0.129412}
  \definecolor{chocolate3}{rgb}{0.803922,0.400000,0.113725}
  \definecolor{chocolate4}{rgb}{0.545098,0.270588,0.074510}
  \definecolor{firebrick1}{rgb}{1.000000,0.188235,0.188235}
  \definecolor{firebrick2}{rgb}{0.933333,0.172549,0.172549}
  \definecolor{firebrick3}{rgb}{0.803922,0.149020,0.149020}
  \definecolor{firebrick4}{rgb}{0.545098,0.101961,0.101961}
  \definecolor{brown1}{rgb}{1.000000,0.250980,0.250980}
  \definecolor{brown2}{rgb}{0.933333,0.231373,0.231373}
  \definecolor{brown3}{rgb}{0.803922,0.200000,0.200000}
  \definecolor{brown4}{rgb}{0.545098,0.137255,0.137255}
  \definecolor{salmon1}{rgb}{1.000000,0.549020,0.411765}
  \definecolor{salmon2}{rgb}{0.933333,0.509804,0.384314}
  \definecolor{salmon3}{rgb}{0.803922,0.439216,0.329412}
  \definecolor{salmon4}{rgb}{0.545098,0.298039,0.223529}
  \definecolor{LightSalmon1}{rgb}{1.000000,0.627451,0.478431}
  \definecolor{LightSalmon2}{rgb}{0.933333,0.584314,0.447059}
  \definecolor{LightSalmon3}{rgb}{0.803922,0.505882,0.384314}
  \definecolor{LightSalmon4}{rgb}{0.545098,0.341176,0.258824}
  \definecolor{orange1}{rgb}{1.000000,0.647059,0.000000}
  \definecolor{orange2}{rgb}{0.933333,0.603922,0.000000}
  \definecolor{orange3}{rgb}{0.803922,0.521569,0.000000}
  \definecolor{orange4}{rgb}{0.545098,0.352941,0.000000}
  \definecolor{DarkOrange1}{rgb}{1.000000,0.498039,0.000000}
  \definecolor{DarkOrange2}{rgb}{0.933333,0.462745,0.000000}
  \definecolor{DarkOrange3}{rgb}{0.803922,0.400000,0.000000}
  \definecolor{DarkOrange4}{rgb}{0.545098,0.270588,0.000000}
  \definecolor{coral1}{rgb}{1.000000,0.447059,0.337255}
  \definecolor{coral2}{rgb}{0.933333,0.415686,0.313726}
  \definecolor{coral3}{rgb}{0.803922,0.356863,0.270588}
  \definecolor{coral4}{rgb}{0.545098,0.243137,0.184314}
  \definecolor{tomato1}{rgb}{1.000000,0.388235,0.278431}
  \definecolor{tomato2}{rgb}{0.933333,0.360784,0.258824}
  \definecolor{tomato3}{rgb}{0.803922,0.309804,0.223529}
  \definecolor{tomato4}{rgb}{0.545098,0.211765,0.149020}
  \definecolor{OrangeRed1}{rgb}{1.000000,0.270588,0.000000}
  \definecolor{OrangeRed2}{rgb}{0.933333,0.250980,0.000000}
  \definecolor{OrangeRed3}{rgb}{0.803922,0.215686,0.000000}
  \definecolor{OrangeRed4}{rgb}{0.545098,0.145098,0.000000}
  \definecolor{red1}{rgb}{1.000000,0.000000,0.000000}
  \definecolor{red2}{rgb}{0.933333,0.000000,0.000000}
  \definecolor{red3}{rgb}{0.803922,0.000000,0.000000}
  \definecolor{red4}{rgb}{0.545098,0.000000,0.000000}
  \definecolor{DeepPink1}{rgb}{1.000000,0.078431,0.576471}
  \definecolor{DeepPink2}{rgb}{0.933333,0.070588,0.537255}
  \definecolor{DeepPink3}{rgb}{0.803922,0.062745,0.462745}
  \definecolor{DeepPink4}{rgb}{0.545098,0.039216,0.313726}
  \definecolor{HotPink1}{rgb}{1.000000,0.431373,0.705882}
  \definecolor{HotPink2}{rgb}{0.933333,0.415686,0.654902}
  \definecolor{HotPink3}{rgb}{0.803922,0.376471,0.564706}
  \definecolor{HotPink4}{rgb}{0.545098,0.227451,0.384314}
  \definecolor{pink1}{rgb}{1.000000,0.709804,0.772549}
  \definecolor{pink2}{rgb}{0.933333,0.662745,0.721569}
  \definecolor{pink3}{rgb}{0.803922,0.568627,0.619608}
  \definecolor{pink4}{rgb}{0.545098,0.388235,0.423529}
  \definecolor{LightPink1}{rgb}{1.000000,0.682353,0.725490}
  \definecolor{LightPink2}{rgb}{0.933333,0.635294,0.678431}
  \definecolor{LightPink3}{rgb}{0.803922,0.549020,0.584314}
  \definecolor{LightPink4}{rgb}{0.545098,0.372549,0.396078}
  \definecolor{PaleVioletRed1}{rgb}{1.000000,0.509804,0.670588}
  \definecolor{PaleVioletRed2}{rgb}{0.933333,0.474510,0.623529}
  \definecolor{PaleVioletRed3}{rgb}{0.803922,0.407843,0.537255}
  \definecolor{PaleVioletRed4}{rgb}{0.545098,0.278431,0.364706}
  \definecolor{maroon1}{rgb}{1.000000,0.203922,0.701961}
  \definecolor{maroon2}{rgb}{0.933333,0.188235,0.654902}
  \definecolor{maroon3}{rgb}{0.803922,0.160784,0.564706}
  \definecolor{maroon4}{rgb}{0.545098,0.109804,0.384314}
  \definecolor{VioletRed1}{rgb}{1.000000,0.243137,0.588235}
  \definecolor{VioletRed2}{rgb}{0.933333,0.227451,0.549020}
  \definecolor{VioletRed3}{rgb}{0.803922,0.196078,0.470588}
  \definecolor{VioletRed4}{rgb}{0.545098,0.133333,0.321569}
  \definecolor{magenta1}{rgb}{1.000000,0.000000,1.000000}
  \definecolor{magenta2}{rgb}{0.933333,0.000000,0.933333}
  \definecolor{magenta3}{rgb}{0.803922,0.000000,0.803922}
  \definecolor{magenta4}{rgb}{0.545098,0.000000,0.545098}
  \definecolor{orchid1}{rgb}{1.000000,0.513726,0.980392}
  \definecolor{orchid2}{rgb}{0.933333,0.478431,0.913725}
  \definecolor{orchid3}{rgb}{0.803922,0.411765,0.788235}
  \definecolor{orchid4}{rgb}{0.545098,0.278431,0.537255}
  \definecolor{plum1}{rgb}{1.000000,0.733333,1.000000}
  \definecolor{plum2}{rgb}{0.933333,0.682353,0.933333}
  \definecolor{plum3}{rgb}{0.803922,0.588235,0.803922}
  \definecolor{plum4}{rgb}{0.545098,0.400000,0.545098}
  \definecolor{MediumOrchid1}{rgb}{0.878431,0.400000,1.000000}
  \definecolor{MediumOrchid2}{rgb}{0.819608,0.372549,0.933333}
  \definecolor{MediumOrchid3}{rgb}{0.705882,0.321569,0.803922}
  \definecolor{MediumOrchid4}{rgb}{0.478431,0.215686,0.545098}
  \definecolor{DarkOrchid1}{rgb}{0.749020,0.243137,1.000000}
  \definecolor{DarkOrchid2}{rgb}{0.698039,0.227451,0.933333}
  \definecolor{DarkOrchid3}{rgb}{0.603922,0.196078,0.803922}
  \definecolor{DarkOrchid4}{rgb}{0.407843,0.133333,0.545098}
  \definecolor{purple1}{rgb}{0.607843,0.188235,1.000000}
  \definecolor{purple2}{rgb}{0.568627,0.172549,0.933333}
  \definecolor{purple3}{rgb}{0.490196,0.149020,0.803922}
  \definecolor{purple4}{rgb}{0.333333,0.101961,0.545098}
  \definecolor{MediumPurple1}{rgb}{0.670588,0.509804,1.000000}
  \definecolor{MediumPurple2}{rgb}{0.623529,0.474510,0.933333}
  \definecolor{MediumPurple3}{rgb}{0.537255,0.407843,0.803922}
  \definecolor{MediumPurple4}{rgb}{0.364706,0.278431,0.545098}
  \definecolor{thistle1}{rgb}{1.000000,0.882353,1.000000}
  \definecolor{thistle2}{rgb}{0.933333,0.823529,0.933333}
  \definecolor{thistle3}{rgb}{0.803922,0.709804,0.803922}
  \definecolor{thistle4}{rgb}{0.545098,0.482353,0.545098}
  \definecolor{gray0}{rgb}{0.000000,0.000000,0.000000}
  \definecolor{grey0}{rgb}{0.000000,0.000000,0.000000}
  \definecolor{gray1}{rgb}{0.011765,0.011765,0.011765}
  \definecolor{grey1}{rgb}{0.011765,0.011765,0.011765}
  \definecolor{gray2}{rgb}{0.019608,0.019608,0.019608}
  \definecolor{grey2}{rgb}{0.019608,0.019608,0.019608}
  \definecolor{gray3}{rgb}{0.031373,0.031373,0.031373}
  \definecolor{grey3}{rgb}{0.031373,0.031373,0.031373}
  \definecolor{gray4}{rgb}{0.039216,0.039216,0.039216}
  \definecolor{grey4}{rgb}{0.039216,0.039216,0.039216}
  \definecolor{gray5}{rgb}{0.050980,0.050980,0.050980}
  \definecolor{grey5}{rgb}{0.050980,0.050980,0.050980}
  \definecolor{gray6}{rgb}{0.058824,0.058824,0.058824}
  \definecolor{grey6}{rgb}{0.058824,0.058824,0.058824}
  \definecolor{gray7}{rgb}{0.070588,0.070588,0.070588}
  \definecolor{grey7}{rgb}{0.070588,0.070588,0.070588}
  \definecolor{gray8}{rgb}{0.078431,0.078431,0.078431}
  \definecolor{grey8}{rgb}{0.078431,0.078431,0.078431}
  \definecolor{gray9}{rgb}{0.090196,0.090196,0.090196}
  \definecolor{grey9}{rgb}{0.090196,0.090196,0.090196}
  \definecolor{gray10}{rgb}{0.101961,0.101961,0.101961}
  \definecolor{grey10}{rgb}{0.101961,0.101961,0.101961}
  \definecolor{gray11}{rgb}{0.109804,0.109804,0.109804}
  \definecolor{grey11}{rgb}{0.109804,0.109804,0.109804}
  \definecolor{gray12}{rgb}{0.121569,0.121569,0.121569}
  \definecolor{grey12}{rgb}{0.121569,0.121569,0.121569}
  \definecolor{gray13}{rgb}{0.129412,0.129412,0.129412}
  \definecolor{grey13}{rgb}{0.129412,0.129412,0.129412}
  \definecolor{gray14}{rgb}{0.141176,0.141176,0.141176}
  \definecolor{grey14}{rgb}{0.141176,0.141176,0.141176}
  \definecolor{gray15}{rgb}{0.149020,0.149020,0.149020}
  \definecolor{grey15}{rgb}{0.149020,0.149020,0.149020}
  \definecolor{gray16}{rgb}{0.160784,0.160784,0.160784}
  \definecolor{grey16}{rgb}{0.160784,0.160784,0.160784}
  \definecolor{gray17}{rgb}{0.168627,0.168627,0.168627}
  \definecolor{grey17}{rgb}{0.168627,0.168627,0.168627}
  \definecolor{gray18}{rgb}{0.180392,0.180392,0.180392}
  \definecolor{grey18}{rgb}{0.180392,0.180392,0.180392}
  \definecolor{gray19}{rgb}{0.188235,0.188235,0.188235}
  \definecolor{grey19}{rgb}{0.188235,0.188235,0.188235}
  \definecolor{gray20}{rgb}{0.200000,0.200000,0.200000}
  \definecolor{grey20}{rgb}{0.200000,0.200000,0.200000}
  \definecolor{gray21}{rgb}{0.211765,0.211765,0.211765}
  \definecolor{grey21}{rgb}{0.211765,0.211765,0.211765}
  \definecolor{gray22}{rgb}{0.219608,0.219608,0.219608}
  \definecolor{grey22}{rgb}{0.219608,0.219608,0.219608}
  \definecolor{gray23}{rgb}{0.231373,0.231373,0.231373}
  \definecolor{grey23}{rgb}{0.231373,0.231373,0.231373}
  \definecolor{gray24}{rgb}{0.239216,0.239216,0.239216}
  \definecolor{grey24}{rgb}{0.239216,0.239216,0.239216}
  \definecolor{gray25}{rgb}{0.250980,0.250980,0.250980}
  \definecolor{grey25}{rgb}{0.250980,0.250980,0.250980}
  \definecolor{gray26}{rgb}{0.258824,0.258824,0.258824}
  \definecolor{grey26}{rgb}{0.258824,0.258824,0.258824}
  \definecolor{gray27}{rgb}{0.270588,0.270588,0.270588}
  \definecolor{grey27}{rgb}{0.270588,0.270588,0.270588}
  \definecolor{gray28}{rgb}{0.278431,0.278431,0.278431}
  \definecolor{grey28}{rgb}{0.278431,0.278431,0.278431}
  \definecolor{gray29}{rgb}{0.290196,0.290196,0.290196}
  \definecolor{grey29}{rgb}{0.290196,0.290196,0.290196}
  \definecolor{gray30}{rgb}{0.301961,0.301961,0.301961}
  \definecolor{grey30}{rgb}{0.301961,0.301961,0.301961}
  \definecolor{gray31}{rgb}{0.309804,0.309804,0.309804}
  \definecolor{grey31}{rgb}{0.309804,0.309804,0.309804}
  \definecolor{gray32}{rgb}{0.321569,0.321569,0.321569}
  \definecolor{grey32}{rgb}{0.321569,0.321569,0.321569}
  \definecolor{gray33}{rgb}{0.329412,0.329412,0.329412}
  \definecolor{grey33}{rgb}{0.329412,0.329412,0.329412}
  \definecolor{gray34}{rgb}{0.341176,0.341176,0.341176}
  \definecolor{grey34}{rgb}{0.341176,0.341176,0.341176}
  \definecolor{gray35}{rgb}{0.349020,0.349020,0.349020}
  \definecolor{grey35}{rgb}{0.349020,0.349020,0.349020}
  \definecolor{gray36}{rgb}{0.360784,0.360784,0.360784}
  \definecolor{grey36}{rgb}{0.360784,0.360784,0.360784}
  \definecolor{gray37}{rgb}{0.368627,0.368627,0.368627}
  \definecolor{grey37}{rgb}{0.368627,0.368627,0.368627}
  \definecolor{gray38}{rgb}{0.380392,0.380392,0.380392}
  \definecolor{grey38}{rgb}{0.380392,0.380392,0.380392}
  \definecolor{gray39}{rgb}{0.388235,0.388235,0.388235}
  \definecolor{grey39}{rgb}{0.388235,0.388235,0.388235}
  \definecolor{gray40}{rgb}{0.400000,0.400000,0.400000}
  \definecolor{grey40}{rgb}{0.400000,0.400000,0.400000}
  \definecolor{gray41}{rgb}{0.411765,0.411765,0.411765}
  \definecolor{grey41}{rgb}{0.411765,0.411765,0.411765}
  \definecolor{gray42}{rgb}{0.419608,0.419608,0.419608}
  \definecolor{grey42}{rgb}{0.419608,0.419608,0.419608}
  \definecolor{gray43}{rgb}{0.431373,0.431373,0.431373}
  \definecolor{grey43}{rgb}{0.431373,0.431373,0.431373}
  \definecolor{gray44}{rgb}{0.439216,0.439216,0.439216}
  \definecolor{grey44}{rgb}{0.439216,0.439216,0.439216}
  \definecolor{gray45}{rgb}{0.450980,0.450980,0.450980}
  \definecolor{grey45}{rgb}{0.450980,0.450980,0.450980}
  \definecolor{gray46}{rgb}{0.458824,0.458824,0.458824}
  \definecolor{grey46}{rgb}{0.458824,0.458824,0.458824}
  \definecolor{gray47}{rgb}{0.470588,0.470588,0.470588}
  \definecolor{grey47}{rgb}{0.470588,0.470588,0.470588}
  \definecolor{gray48}{rgb}{0.478431,0.478431,0.478431}
  \definecolor{grey48}{rgb}{0.478431,0.478431,0.478431}
  \definecolor{gray49}{rgb}{0.490196,0.490196,0.490196}
  \definecolor{grey49}{rgb}{0.490196,0.490196,0.490196}
  \definecolor{gray50}{rgb}{0.498039,0.498039,0.498039}
  \definecolor{grey50}{rgb}{0.498039,0.498039,0.498039}
  \definecolor{gray51}{rgb}{0.509804,0.509804,0.509804}
  \definecolor{grey51}{rgb}{0.509804,0.509804,0.509804}
  \definecolor{gray52}{rgb}{0.521569,0.521569,0.521569}
  \definecolor{grey52}{rgb}{0.521569,0.521569,0.521569}
  \definecolor{gray53}{rgb}{0.529412,0.529412,0.529412}
  \definecolor{grey53}{rgb}{0.529412,0.529412,0.529412}
  \definecolor{gray54}{rgb}{0.541176,0.541176,0.541176}
  \definecolor{grey54}{rgb}{0.541176,0.541176,0.541176}
  \definecolor{gray55}{rgb}{0.549020,0.549020,0.549020}
  \definecolor{grey55}{rgb}{0.549020,0.549020,0.549020}
  \definecolor{gray56}{rgb}{0.560784,0.560784,0.560784}
  \definecolor{grey56}{rgb}{0.560784,0.560784,0.560784}
  \definecolor{gray57}{rgb}{0.568627,0.568627,0.568627}
  \definecolor{grey57}{rgb}{0.568627,0.568627,0.568627}
  \definecolor{gray58}{rgb}{0.580392,0.580392,0.580392}
  \definecolor{grey58}{rgb}{0.580392,0.580392,0.580392}
  \definecolor{gray59}{rgb}{0.588235,0.588235,0.588235}
  \definecolor{grey59}{rgb}{0.588235,0.588235,0.588235}
  \definecolor{gray60}{rgb}{0.600000,0.600000,0.600000}
  \definecolor{grey60}{rgb}{0.600000,0.600000,0.600000}
  \definecolor{gray61}{rgb}{0.611765,0.611765,0.611765}
  \definecolor{grey61}{rgb}{0.611765,0.611765,0.611765}
  \definecolor{gray62}{rgb}{0.619608,0.619608,0.619608}
  \definecolor{grey62}{rgb}{0.619608,0.619608,0.619608}
  \definecolor{gray63}{rgb}{0.631373,0.631373,0.631373}
  \definecolor{grey63}{rgb}{0.631373,0.631373,0.631373}
  \definecolor{gray64}{rgb}{0.639216,0.639216,0.639216}
  \definecolor{grey64}{rgb}{0.639216,0.639216,0.639216}
  \definecolor{gray65}{rgb}{0.650980,0.650980,0.650980}
  \definecolor{grey65}{rgb}{0.650980,0.650980,0.650980}
  \definecolor{gray66}{rgb}{0.658824,0.658824,0.658824}
  \definecolor{grey66}{rgb}{0.658824,0.658824,0.658824}
  \definecolor{gray67}{rgb}{0.670588,0.670588,0.670588}
  \definecolor{grey67}{rgb}{0.670588,0.670588,0.670588}
  \definecolor{gray68}{rgb}{0.678431,0.678431,0.678431}
  \definecolor{grey68}{rgb}{0.678431,0.678431,0.678431}
  \definecolor{gray69}{rgb}{0.690196,0.690196,0.690196}
  \definecolor{grey69}{rgb}{0.690196,0.690196,0.690196}
  \definecolor{gray70}{rgb}{0.701961,0.701961,0.701961}
  \definecolor{grey70}{rgb}{0.701961,0.701961,0.701961}
  \definecolor{gray71}{rgb}{0.709804,0.709804,0.709804}
  \definecolor{grey71}{rgb}{0.709804,0.709804,0.709804}
  \definecolor{gray72}{rgb}{0.721569,0.721569,0.721569}
  \definecolor{grey72}{rgb}{0.721569,0.721569,0.721569}
  \definecolor{gray73}{rgb}{0.729412,0.729412,0.729412}
  \definecolor{grey73}{rgb}{0.729412,0.729412,0.729412}
  \definecolor{gray74}{rgb}{0.741176,0.741176,0.741176}
  \definecolor{grey74}{rgb}{0.741176,0.741176,0.741176}
  \definecolor{gray75}{rgb}{0.749020,0.749020,0.749020}
  \definecolor{grey75}{rgb}{0.749020,0.749020,0.749020}
  \definecolor{gray76}{rgb}{0.760784,0.760784,0.760784}
  \definecolor{grey76}{rgb}{0.760784,0.760784,0.760784}
  \definecolor{gray77}{rgb}{0.768627,0.768627,0.768627}
  \definecolor{grey77}{rgb}{0.768627,0.768627,0.768627}
  \definecolor{gray78}{rgb}{0.780392,0.780392,0.780392}
  \definecolor{grey78}{rgb}{0.780392,0.780392,0.780392}
  \definecolor{gray79}{rgb}{0.788235,0.788235,0.788235}
  \definecolor{grey79}{rgb}{0.788235,0.788235,0.788235}
  \definecolor{gray80}{rgb}{0.800000,0.800000,0.800000}
  \definecolor{grey80}{rgb}{0.800000,0.800000,0.800000}
  \definecolor{gray81}{rgb}{0.811765,0.811765,0.811765}
  \definecolor{grey81}{rgb}{0.811765,0.811765,0.811765}
  \definecolor{gray82}{rgb}{0.819608,0.819608,0.819608}
  \definecolor{grey82}{rgb}{0.819608,0.819608,0.819608}
  \definecolor{gray83}{rgb}{0.831373,0.831373,0.831373}
  \definecolor{grey83}{rgb}{0.831373,0.831373,0.831373}
  \definecolor{gray84}{rgb}{0.839216,0.839216,0.839216}
  \definecolor{grey84}{rgb}{0.839216,0.839216,0.839216}
  \definecolor{gray85}{rgb}{0.850980,0.850980,0.850980}
  \definecolor{grey85}{rgb}{0.850980,0.850980,0.850980}
  \definecolor{gray86}{rgb}{0.858824,0.858824,0.858824}
  \definecolor{grey86}{rgb}{0.858824,0.858824,0.858824}
  \definecolor{gray87}{rgb}{0.870588,0.870588,0.870588}
  \definecolor{grey87}{rgb}{0.870588,0.870588,0.870588}
  \definecolor{gray88}{rgb}{0.878431,0.878431,0.878431}
  \definecolor{grey88}{rgb}{0.878431,0.878431,0.878431}
  \definecolor{gray89}{rgb}{0.890196,0.890196,0.890196}
  \definecolor{grey89}{rgb}{0.890196,0.890196,0.890196}
  \definecolor{gray90}{rgb}{0.898039,0.898039,0.898039}
  \definecolor{grey90}{rgb}{0.898039,0.898039,0.898039}
  \definecolor{gray91}{rgb}{0.909804,0.909804,0.909804}
  \definecolor{grey91}{rgb}{0.909804,0.909804,0.909804}
  \definecolor{gray92}{rgb}{0.921569,0.921569,0.921569}
  \definecolor{grey92}{rgb}{0.921569,0.921569,0.921569}
  \definecolor{gray93}{rgb}{0.929412,0.929412,0.929412}
  \definecolor{grey93}{rgb}{0.929412,0.929412,0.929412}
  \definecolor{gray94}{rgb}{0.941176,0.941176,0.941176}
  \definecolor{grey94}{rgb}{0.941176,0.941176,0.941176}
  \definecolor{gray95}{rgb}{0.949020,0.949020,0.949020}
  \definecolor{grey95}{rgb}{0.949020,0.949020,0.949020}
  \definecolor{gray96}{rgb}{0.960784,0.960784,0.960784}
  \definecolor{grey96}{rgb}{0.960784,0.960784,0.960784}
  \definecolor{gray97}{rgb}{0.968627,0.968627,0.968627}
  \definecolor{grey97}{rgb}{0.968627,0.968627,0.968627}
  \definecolor{gray98}{rgb}{0.980392,0.980392,0.980392}
  \definecolor{grey98}{rgb}{0.980392,0.980392,0.980392}
  \definecolor{gray99}{rgb}{0.988235,0.988235,0.988235}
  \definecolor{grey99}{rgb}{0.988235,0.988235,0.988235}
  \definecolor{gray100}{rgb}{1.000000,1.000000,1.000000}
  \definecolor{grey100}{rgb}{1.000000,1.000000,1.000000}
  \definecolor{dark grey}{rgb}{0.662745,0.662745,0.662745}
  \definecolor{DarkGrey}{rgb}{0.662745,0.662745,0.662745}
  \definecolor{dark gray}{rgb}{0.662745,0.662745,0.662745}
  \definecolor{DarkGray}{rgb}{0.662745,0.662745,0.662745}
  \definecolor{dark blue}{rgb}{0.000000,0.000000,0.545098}
  \definecolor{DarkBlue}{rgb}{0.000000,0.000000,0.545098}
  \definecolor{dark cyan}{rgb}{0.000000,0.545098,0.545098}
  \definecolor{DarkCyan}{rgb}{0.000000,0.545098,0.545098}
  \definecolor{dark magenta}{rgb}{0.545098,0.000000,0.545098}
  \definecolor{DarkMagenta}{rgb}{0.545098,0.000000,0.545098}
  \definecolor{dark red}{rgb}{0.545098,0.000000,0.000000}
  \definecolor{DarkRed}{rgb}{0.545098,0.000000,0.000000}
  \definecolor{light green}{rgb}{0.564706,0.933333,0.564706}
  \definecolor{LightGreen}{rgb}{0.564706,0.933333,0.564706}

\usepackage{graphicx}
\usepackage{color}

\usepackage[T1]{fontenc}
\usepackage{aecompl}

\newcommand{\Msun}{M$_{\odot}$}

\begin{document}

\title[Off The Beaten Path]{Off the Beaten Path: A New Approach to Realistically Model The Orbital Decay of Supermassive Black Holes in Galaxy Formation Simulations}
\author[M. Tremmel et al.]{M. ~Tremmel$^{1}$\thanks{email: mjt29@uw.edu}, F. ~Governato$^{1}$, M. ~Volonteri$^{2,3}$, T. ~R. ~Quinn$^{1}$\\
$^1$Astronomy Department, University of Washington, Box 351580, Seattle, WA, 98195-1580\\
$^2$Institut dÕAstrophysique, UMR 7095 CNRS, Universit«e Pierre et Marie Curie, 98bis Blvd Arago, 75014 Paris, France\\
$^3$ Department of Astronomy, University of Michigan, Ann Arbor, MI, USA}

\pagerange{\pageref{firstpage}--\pageref{lastpage}} \pubyear{2015}

\maketitle

\label{firstpage}

\begin{abstract}

We introduce a sub-grid force correction term to better model the dynamical friction (DF) experienced by a supermassive black hole (SMBH) as it orbits within its host galaxy.  This new approach accurately follows a SMBH's orbital decay and drastically improves over commonly used `advection' methods. The force correction introduced here naturally scales with the force resolution of the simulation and converges as resolution is increased. In controlled experiments we show how the orbital decay of the SMBH closely follows analytical predictions when particle masses  are significantly smaller than that of the SMBH. In a cosmological simulation of the assembly of a small galaxy, we show how our method allows for realistic black hole orbits. This approach overcomes the limitations of the advection scheme, where black holes are rapidly and artificially pushed toward the halo center and then forced to merge, regardless of their orbits.  We find that SMBHs from merging dwarf galaxies can spend significant time away from the center of the remnant galaxy. Improving the modeling of SMBH orbital decay will help in making robust predictions of the growth, detectability, and merger rates of SMBHs, especially at low galaxy masses or at high redshift.
\end{abstract}


\maketitle

\section{Introduction}
Supermassive Black Holes (SMBHs) represent a crucial, though still poorly understood, aspect of galaxy evolution theory. SMBHs with as much as $10^9$ M$_{\odot}$ in mass power luminous quasars observed at z $>$ 6 \citep{fan01, mortlock11}, while in the local Universe, SMBHs are ubiquitous in both massive galaxies \citep[e.g.][]{Gehren84,kormendy95,kormendy2013} and small, bulge-less disk galaxies \citep{shields08,Filippenko03} as well as dwarfs \citep{reines11,reines12,reines13,moran14}. Empirical scaling relations between the mass of SMBHs and the stellar mass and velocity dispersion of their host galaxies are indicative of co-evolution \citep{haring04,gultekin09,schramm2013,kormendy2013}. Powerful outflows resulting from feedback from an active nucleus have been observed \citep{feruglio10,alatalo11} and are thought to play a critical role in the quenching of star formation in massive galaxies and the exponential cutoff of the galaxy luminosity function at high masses \citep{springel05b,bower06,croton09,Teyssier11}.

Numerical simulations have proven to be a critical tool for understanding the formation and evolution of galaxies and extensive work has already been done to implement SMBHs into these simulations \citep[e.g.][]{dimatteoBH03,hopkins05,sijackiBH2009}. Due to the scale of these simulations, which often include a full cosmological environment, the resolution is necessarily limited. Accretion onto SMBHs and the following feedback processes are hence implemented via sub-grid prescriptions, under the broad assumption that conditions at the smallest resolved scale drive the BH evolution at the scale of just a few parsecs. One major obstacle is that it becomes necessary to accurately follow the dynamics of a single object (the SMBH), something that cosmological simulations with force resolution larger than a few pc are inherently not well equipped to do. These dynamics can have important consequences for the growth of SMBHs, particularly during galaxy mergers. Detailed simulations of galaxy mergers show that the dynamics of SMBHs are intimately connected with accretion and vice versa \citep[e.g.][]{callegari11}.

In this paper, we describe our approach to take into account unresolved dynamical friction (DF) in the orbital evolution of SMBHs and show that our prescription is a promising and realistic alternative to advection that can easily be implemented in existing codes. We show that this approach, when applied to simulations with the typical resolution of modern cosmological simulations ($\sim0.1-0.5$~kpc) leads to more realistic dynamics that match well with analytic approximations for DF timescales. In Section 2 we summarize some current methods for correcting SMBH dynamics in cosmological simulations and why a new method is needed. In section 3 we present our method. In section 4 we present our results, comparing our method with a popular advection prescription in both an isolated dark matter (DM) halo and a fully cosmological zoomed-in simulation of a dwarf galaxy. In section 5 we summarize and discuss our results. 

\section{The Need for a New Model}
Dynamical friction, the force exerted by the gravitational wake
caused by a massive object moving in an extended medium
\citep{DF43,BinneyTremaine} causes the orbits of SMBHs to decay towards
the center of massive galaxies \citep{G94,stelios05}.  In cosmological
simulations, modeling DF is particulary challenging,
as the particle representing a SMBH is often only a few times the mass
of the background particles. In this case SMBH orbits can become
dynamically heated due to the limited mass resolution and the
resulting spurious collisionality from a noisy gravitational
potential \citep{collisional90}. This numerical heating can cause the
black holes to gain/lose energy and be unrealistically perturbed away
from the center of its host halo.  In order to lessen collisionality
and improve performance, N-body simulations employ a gravitational
softening length, a characteristic scale below which the
gravitational force between two particles becomes damped. This
mechanism, while necessary numerically, hinders DF by
preventing the close interactions that are necessary to form a wake in
the vicinity of the SMBH.

To account for the unrealistic dynamics of SMBHs in cosmological simulations, many groups employ artificial advection schemes that change the motion of SMBHs. Each method comes with its own drawbacks. For example, placing the SMBH at the position of the lowest potential gas particle around it \citep[e.g.][]{springel05,BoothBH2009} causes chaotic motions, especially when relative velocity constraints are put on the gas particles \citep{wurster13a}. Utilizing a large `tracer mass' with which the SMBH gravitationally interacts with its surroundings \citep{debuhr14} affects the morphology in the galactic center as well as the accretion history of the SMBH \citep{wurster13a}. Pushing the SMBH a certain distance in the direction of either the local stellar density gradient \citep{Okamoto2008} or the local center of mass \citep{wurster13a} can avoid chaotic motions but adds an additional free parameter, the distance the SMBH is pushed each step. Even disregarding their individual drawbacks, none of these solutions are ideal, as they all require the explicit assumption that SMBHs rapidly decay into and remain stable at the center of their host galaxies , which is often not accurate \citep{bellovaryBH10}. 

In strongly interacting systems, where
SMBHs are thought to go through much of their growth and possibly their formation \citep{mayer07}, the inner
regions of a galaxy are likely to experience strong potential
fluctuations, affecting the dynamics of the SMBH. This
 could be especially evident in dwarf galaxies, where black
holes exist \citep{reines13,moran14} within a shallow potential with a cored
density profile \citep{oh11,Teyssier13}, making perturbations to their orbits
more likely. Additionally, SMBHs accreted during both major and minor
mergers, especially dry mergers \citep{stelios05}, do not immediately 
sink to the center of the galaxy, creating a population of small `wandering' SMBHs
\citep{Islam03,volonteriOffCenBH05,bellovaryBH10}. Advection, by assuming that orbital decay
timescales are always short, could then artificially increase the mass
of central SMBHs, through inflated merger and accretion rates.

More advanced approaches have been explored. For example, \citet{dubois2013} include a sub-grid prescription for gas dynamical friction acting on the SMBH. While this is a promising method, it requires assumptions about the multiphase nature and equation of state of the gas far below the resolution limit of any cosmological simulation ($\sim 1-5$ pc). \citet{lupi15} increase the resolution around SMBHs to attain very accurate dynamics, but their method is applicable only for very high resolution simulations meant to closely follow BH-BH mergers, not cosmological simulations.

The method we propose in this Letter is a simple solution to correcting SMBH dynamics in cosmological simulations. Our approach is to estimate the unresolved dynamical friction felt by the SMBH and apply the appropriate force to the SMBH particle. This is a significant improvement over artificial advection, as it makes no assumption about where the SMBHs should be located in a galaxy at a given time and it has no explicit effect on the SMBH surroundings. In addition, it requires minimal assumptions about the state of the simulation below the resolution limit and it naturally converges with increasing resolution.


\section{Methodology}
\subsection{The Test Simulations}
To test the dynamics of black holes in a realistic DM halo,
we generate initial conditions for a collapsing overdensity following
the procedure of \citet{evrard88}\footnote[1]{Created using the package
  ICInG created by M. Tremmel. The code is publicly available and can be downloaded at
  https://github.com/mtremmel/ICInG.git}. We begin the simulation by approximating the state of a halo beginning to collapse at high redshift. We then allow the halo to collapse before implanting a black hole of mass $10^6$ M$_{\odot}$. We run a suite of simulations at different
mass and spatial resolutions (see Table 1) with a massive `black hole'
particle initially 1) at the center while the halo is still actively collapsing or 2) on an eccentric orbit after the halo has mostly relaxed. The
resolution of the `Low Res' and `Oversampled' runs are comparable to
current uniform volume simulations
\citep{keres12,illustrisBH14,eagle14a,dubois14}, while the `High Res'
simulation is representative of high resolution `zoomed-in'
simulations \citep{G12,christensen14,hopkins14}. The
  force softening lengths adopted are typical of cosmological runs, while in
  `Small Soft', a variant of the `High Res' simulations, the force
  resolution is only 10pc, typical of simulations of isolated binary
  mergers \citep{capelo15} or very high-res cosmological zoomed-in simulations \citep[e.g.][]{dubois14II}.

 We verified that the
halo formed in the collapse has a density profile typical of CDM halos
\citep{navarro96}.  The halo we use here has (after a time consistent with z $\sim 0$) M$_{vir}\sim
2\times10^{11}$ M$_{\odot}$, R$_{vir}\sim115$ kpc, and a concentration
$c\sim 4.5$.  An analytic expression of the approximate timescale for
a rigid object to sink to the center of a DM halo was
calculated by \citet{taffoni03}. Given these conditions and the mass
of the test black hole, the estimated DF timescale
($\tau_{DF}$) for an eccentric (v$=0.1 $v$_{circ}$) orbit at an initial
distance of $2$ kpc from the halo center is approximately $1.8$
Gyr. This initial orbit was chosen as typical of a SMBH after its
parent satellite has been tidally disrupted. It has a non-negligible
$\tau_{DF}$, but still much less than a Hubble time.

\begin{table}
\caption{Information about the resolution of each isolated dark matter halo test simulation.}
\label{symbols}
\begin{tabular}{@{}cccc}
\hline
Name & Number & Dark Matter & Softening \\
            & of Particles & Mass (M$_{\odot}^{\ast}$ & length, $\epsilon_g$ (pc)$^{\dagger}$ \\
\hline
Low Res & $1.05\times10^6$ & $9.78 \times 10^5$ & 311\\
Oversampled & $8.39\times10^6$ & $1.22  \times 10^5$ & 311\\
High Res & $6.71\times10^7$ & $1.53 \times 10^4$ & 77\\
Small Soft & $6.71\times10^7$  & $1.53 \times 10^4 $&10\\
\hline
\end{tabular}

\medskip
$\ast$ Mass of the dark matter particles in the simulation\\
$\dagger$ spline kernel gravitational softening
\end{table}

\begin{figure*}
\centering
\includegraphics[scale=0.52]{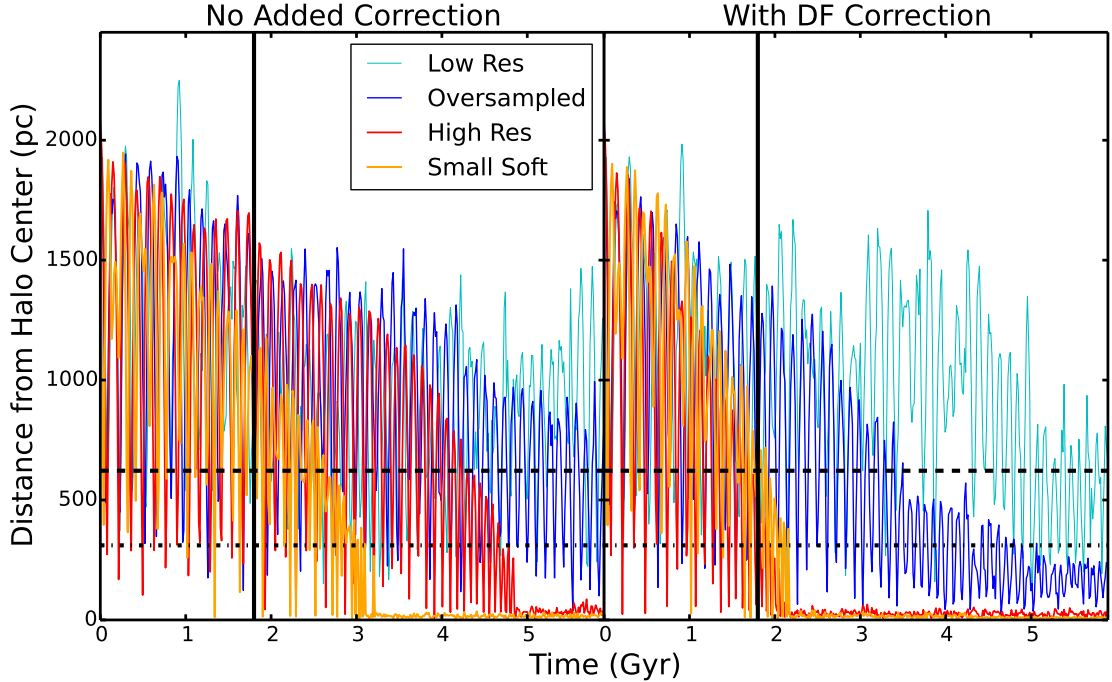}
\caption{{\sc{Effects of the DF correction: }}Distance of a test $10^6$ M$_{\odot}$ black hole from halo
  center as a function of time at different resolutions. Dashed-dot
  and dashed lines indicate $\epsilon_g$ and $2\epsilon_g$
  respectively for both the Low Res and Oversampled
  models. The black hole starts on an eccentric (v$=0.1$ v$_{circ}$) orbit with apocenter
  of $2$ kpc.  The vertical solid line represents the analytically
  derived timescale for orbital decay. When the DF correction is
  applied, marked improvement is seen for all models except `Low
  Res', which experiences too much dynamical heating due to the lower mass resolution.
    The orbits of `High Res' and `Small Soft' are very nearly
  the same once the correction is implemented, indicating numerical
  convergence when the DM particle mass is $\sim$ 10$^4$ \Msun.}
  \end{figure*}
  

\subsection{The Dynamical Friction Prescription}
Dynamical friction occurs due to both large \citep{colpi99} and small scale perturbations to the black hole's surroundings. We consider perturbations on scales larger than the gravitational softening length, $\epsilon_{g}$, to be well resolved. On these scales the potential should be smooth, so long as enough particles are used so that dynamical heating is minimized.

The acceleration a black hole of mass M$_{BH}$ feels from DF due to particles of mass  m$_a$ $\ll$  M$_{BH}$  is given by the following formula:

\begin{equation}
\mathbf{a}_{DF} = -4\pi\mathrm{G}^2\mathrm{M}_{BH}\mathrm{m}_a\mathrm{ln}\Lambda\int d^3\textbf{v}_a f(\textbf{v}_a)\frac{\textbf{v}_{BH} - \textbf{v}_a}{|\textbf{v}_{BH} - \textbf{v}_a|^3}
\end{equation}

Where $f()$ is the velocity distribution function, $\mathrm{ln}\Lambda$ is the Coulomb logarithm, $\textbf{v}_a$ is the velocity of the surrounding background objects relative to the local center of mass (COM) velocity, and $\textbf{v}_{BH}$ is the relative velocity of the black hole. We assume that within $\epsilon_{g}$ from the black hole the velocity distribution is isotropic, giving Chandrasekhar's Dynamical Friction Formula \citep{DF43}.  

\begin{equation}
\mathbf{a}_{DF} = -4\pi\mathrm{G}^2\mathrm{M}\mathrm{m}_a\mathrm{ln}\Lambda\frac{\mathbf{v}_{BH}}{\mathrm{v}_{BH}^3}\int_{0}^{\mathrm{v_{BH}}} d\mathrm{v}_a\mathrm{v}_a^2 f(\textbf{v}_a)
\end{equation}

This can be further simplified by substituting the integral for $\rho(<\mathrm{v}_{BH})$, which is the density of particles moving slower than the black hole.

\begin{equation}
\mathbf{a}_{DF} = -4\pi\mathrm{G}^2\mathrm{M}\rho(<\mathrm{v}_{BH})\mathrm{ln}\Lambda\frac{\mathbf{v}_{BH}}{\mathrm{v}_{BH}^3}
\end{equation}

The Coulomb logarithm depends on the maximum and minimum impact parameters, b$_{max}$ and b$_{min}$, such that $\mathrm{ln}\Lambda \sim \mathrm{ln}(\frac{\mathrm{b}_{max}}{\mathrm{b}_{min}})$. Because DF is well resolved at scales greater than the softening length, we set b$_{max}$ = $\epsilon_{g}$ to avoid double counting frictional forces that are already occurring. For the minimum impact parameter, we take it to be the minimum $90^{\circ}$ deflection radius, with a lower limit set to the Schwarzschild Radius, $\mathrm{R}_{Sch}$.

\begin{equation}
\mathrm{b}_{min} = max(\mathrm{b}_{90},\mathrm{R}_{Sch}); \mathrm{b}_{90} = \frac{\mathrm{GM}_{BH}}{\mathrm{v}_{BH}^2}
\end{equation}

For the calculation, we use 64 collisionless particles (i.e. dark
matter and star particles, if present) closest to the black hole.  We
calculate the velocity of each particle relative to the COM
velocity of those 64 particles. We verified that our results do not depend strongly on the 
number of neighbors used, although using too few particles could result in numerical noise in the calculation of this force.
Since we are explicitly assuming the velocity distribution is
isotropic, the following must be true.

\begin{equation}
\rho(<\mathrm{v}_{BH}) = \frac{\mathrm{M}(<\mathrm{v}_{BH})}{\mathrm{M}_{\mathrm{total}}} \rho
\end{equation}

Where $\rho$, the total density around the black hole, is calculated by smoothing over the chosen 64 particles, i.e.  $\rho = \sum_{i}^{64} m_i W(\mathbf{r}_{BH} - \mathbf{r}_i,h)$. $\mathrm{M}(<\mathrm{v}_{BH})$ is the total mass of the chosen particles that are moving slower than the black hole relative to the local center of mass velocity and $\mathrm{M}_{\mathrm{total}}$ is the summed mass of all 64 particles.

The resulting acceleration (from eq. 3) is added to the black hole's
current acceleration, to be integrated the following time step. As the
spatial resolution or black hole mass increases (or the velocity of the black hole
decreases) $\mathrm{b}_{min}$ will become greater than
$\mathrm{b}_{max}$, in which case we claim DF is being
fully resolved and therefore the correction is not
needed.  

This method is not accounting for DF from gas, which can have important effects for supersonic black holes in regions where gas density dominates stars and dark matter  \citep{ostriker99,chapon13}. This should not occur often on the scales relevant in these simulations. The center of larger galaxies are dominated by stars  and smaller galaxies have significant star and dark matter fractions within the central regions \citep{oh08,oh11}, so this effect will only be a minor correction in most cases. DF may be overestimated within resonant DM cores where DF can become much less efficient \citep{reed06}. This effect is secondary and mainly important when the gravitational softening length is appreciable compared to the size of the core structure. Often the orbital differences should be smaller than or similar to the resolution limit and therefore unimportant. Additionally, interactions with clumpy gas has been shown to significantly increase the timescale for the orbital decay of SMBH binaries below $\sim 100 pc$ \citep{fiacconi13,roskar15}.  However, this effect would not be well resolved by even high resolution `zoomed-in' cosmological simulations.


  \begin{figure}
\centering
\includegraphics[scale=0.4]{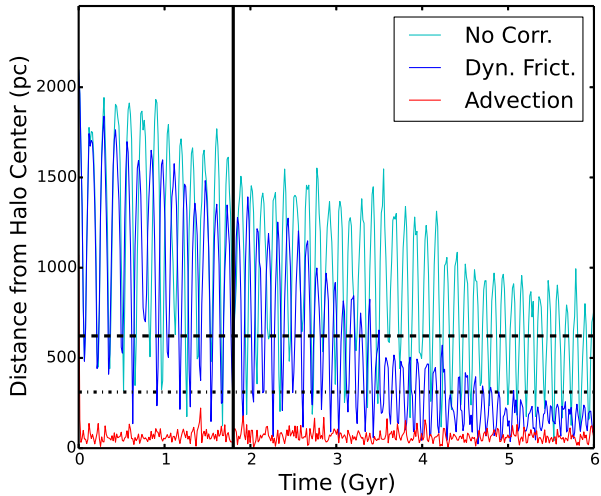}
\caption{{\sc{DF correction vs Advection: }} Results from the `Oversampled/ model when implementing our DF
  correction (blue) compared with a commonly used advection routine
  (red) and no correction to dynamics (cyan). Dashed-dot and dashed lines indicate $\epsilon_g$ and $2\epsilon_g$
  respectively.The black hole starts on an eccentric (v$=0.1$ v$_{circ}$) orbit with apocenter
  of $2$ kpc. Advection immediately pushes the off-center black hole to the
  center, missing the orbital decay that our method captures well. Without any correction, the orbit decays far too slowly, remaining far ($> 2\epsilon_g$) from halo center even after 6 Gyr.}
 \end{figure}

\section{Results}

\subsection{Isolated Dark Matter Halo}
We find that, for SMBHs placed in the center of a collapsing halo, only the Low Res simulation experiences significant dynamical heating, causing the BH to be unrealistically perturbed away far from halo center.  In the rest of the runs the BH remains within one softening length from halo center for 6 Gyrs and shows no sign of heating. This is due to the higher mass resolution present in those runs. Simulations including either our DF correction or an advection correction (see below) have the same results, showing little difference from simulations with no correction at all for this scenario.


Figure 1 shows the orbital evolution of a
black hole initially on an eccentric (v$=0.1$ v$_{circ}$) orbit with
apocenter of $2$ kpc, placed in the halo after it has finished most of its collapse. The center of the halo is defined at each step using the shrinking spheres method \citep{power03}. We verified that the density maximum and potential minimum coincide within much less than the force resolution. The vertical line represents the dynamical friction timescale for the orbit derived from the analytic model of \citet{taffoni03} and the horizontal lines represent $\epsilon_{g}$ and $2\epsilon_{g}$. Without the DF correction, only the
`Small Soft' model, with 10 pc spatial resolution and DM particle mass almost 100 times smaller than the BH, is able to show substantial orbital decay within $2\tau_{DF}$.  Implementing our DF correction results in a noticeable improvement for the orbital decay, even at the relatively modest resolution of the Oversampled model, where it falls to within $2\epsilon_{g}$ of the center before $2\tau_{DF}$. At higher resolution, the dynamics converge to closely match with the analytical approximation. Note that even for our highest resolution simulation, Small Soft, the DF correction causes the SMBH to sink almost 1 Gyr sooner. These are very encouraging results, as they indicate that this correction results in realistic black hole orbital evolution even at resolutions attainable in large volume simulations and it has important consequences even at the highest resolutions tested here.

 Figure 2 compares the performance of our DF prescription to that of a
 commonly used advection scheme used in \citet{sijacki2007} and
 various other simulations. The test is done using the Oversampled
 run, as this most closely resembles the resolution of a cosmological volume simulation.
 The BH is placed on an eccentric orbit, as in Figure 1.
  The advection scheme adopted repositions the black hole each
 time step to the position of the lowest potential particle within its
 32 nearest neighbors while keeping the velocity unchanged. Not
 surprisingly, this results in the black hole staying very close to
 the center of the halo even when initially set on an off-center orbit.  The DF correction captures the more gradual orbital decay that the advection scheme completely misses. The run with no dynamical correction fails to have the BH sink within $2\epsilon_g$ even after 6 Gyr.

This is an important conclusion because these different orbital
evolutions would result in drastically different accretion histories
for the black hole. Off center BHs should accrete less due to lower gas densities. 
Additionally, simulations that utilize advection
would have black holes merge much sooner than what
is predicted by their orbital decay timescale. Our improved method should then
have important implications for the growth and merger rate of SMBHs in cosmological
simulations of galaxy formation.

\begin{figure}
\centering
\includegraphics[trim=0.55in 0.05in 1.1in 0.7in,clip,scale=0.335]{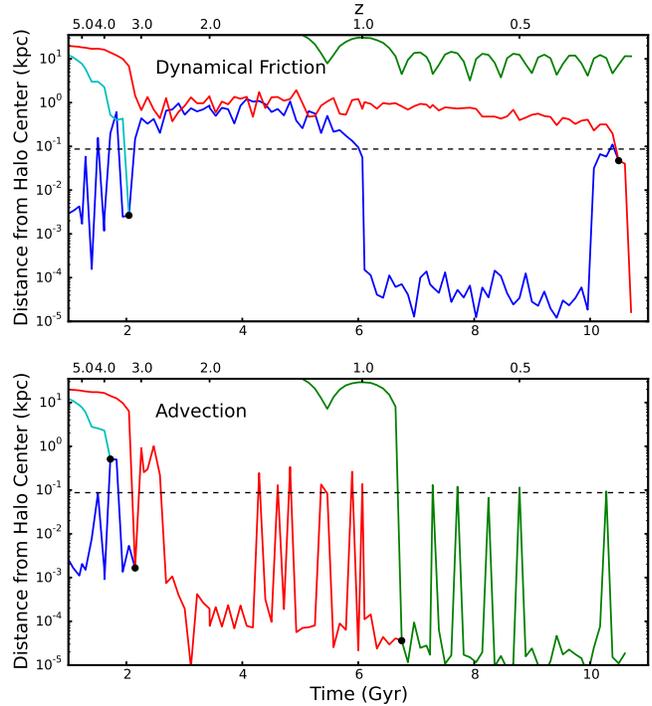}
\caption{{\sc{DF correction in a cosmological dwarf simulation:}}The dynamics of four black holes in the cosmological zoomed-in dwarf galaxy simulation with DF (top) and advection (bottom). These are the black holes that end up in the most massive system by the end of the simulation. Each colored line traces the distance of a black hole from the center of the most massive halo. Black dots mark merger events and the dashed lines mark the gravitational softening length of the simulation (87 pc). Which of the two black holes emerges from a merger event and which is `eaten' is unimportant. DF is able to sustain a long-lived dual black hole system (blue and red) while the advection scheme causes them to quickly merge. The green black hole remains on a very wide orbit in the DF run, but is quickly and unrealistically pulled to the center with advection.}
\end{figure}

\begin{figure*}
\centering
\includegraphics[scale=0.48]{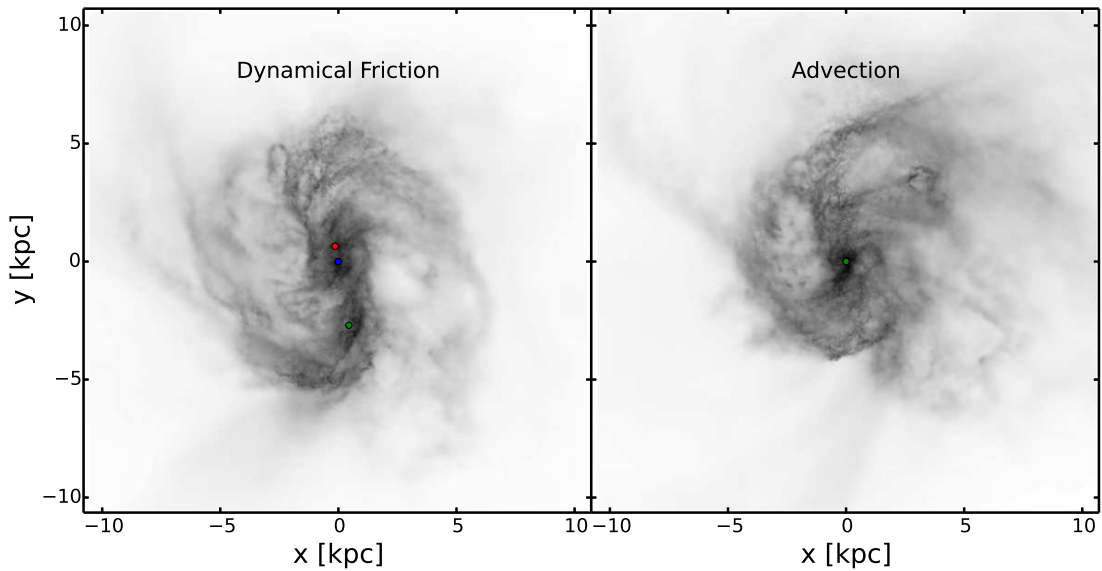}
\caption{{\sc BHs in a Cosmological Dwarf:}
A snapshot of a zoomed-in cosmological simulation of a forming dwarf galaxy at z $= 0.846$. The gas density integrated along the line of sight is shown with darker colors indicating higher densities. In the dynamical friction simulation, a previous  merger has created a central binary BH system (red and blue, see Figure 3). The separation of $\sim1$ kpc is well resolved by the simulation, which has a force resolution of 87 pc. A more recent merger has set a third BH (in green) on a wide orbit (see Figure 3). In the image, the green BH is at its closest approach to galactic center. In the advection simulation, all BHs are quickly pushed to the center, where they merge, causing the simulation to miss these more realistic BH orbits.}
\end{figure*}
 
\subsection{Cosmological Dwarf Galaxy Simulation}
As a first test of the dynamics of SMBHs in a fully cosmological
 setting, we  run a high resolution `zoomed-in' simulation that results in two 
  dwarf galaxies with masses $\sim10^{10}$ M$_{\odot}$ at z = 0. The simulation has a
resolution similar to our High Res isolated halo model, with dark
matter particle mass $1.6 \times 10^4$ M$_{\odot}$, gas particle mass
$3.3 \times 10^3$ M$_{\odot}$, and gravitational softening of only
$87$ pc. We showed in the previous section that at this resolution our DF prescription gives results
that match analytic models. We chose a dwarf galaxy for this test because SMBHs are more likely to become perturbed away from 
galactic center, given their shallow gravitational potential and actively evolving, cored DM profile \citep{G12,PG13}. This 
will guarantee a useful test environment for exploring the differences between out method and advection.

This test is also topical, as there is a growing sample of dwarf galaxies with detected SMBHs \citep{reines13,moran14}.  Realistic numerical studies of BH formation and growth in these small galaxies, focusing on their occupation fraction and how they and their host galaxies evolve toward the correlation with the stellar velocity dispersion, would provide vital constraints on BH seed masses and early growth mechanisms \citep{volonteri08,volonteri2009,volonteri2010AARV}.

 We use the new N-body + SPH code {\sc ChaNGa} \citep{changa15}, which includes all of the physics modules previously implemented in {\sc Gasoline} \citep{wadsley04} such as
hydrodynamics, gas cooling, a cosmic UV background, star formation and
SNe feedback. The `zoomed-in' approach preserves the large scale
tidal field while allowing us to model a small region at high resolution. This is similar to the simulation described in \citet{G10} and
in \citet{shen14}.  In this particular simulation, stars and DM densities dominate that of gas within the inner regions by more than a factor of $10$. Examining other dwarf galaxy simulations \citep[e.g.][]{G10}, we find that gas densities can often reach similar values to DM and stars. In either case the contribution from gas to DF is only a minor correction, at most a factor of a few, and negligible in our current example. 

By z $<$1 these systems are a good representation
of real dwarf galaxies, with an extended stellar disk, no bulge, a
high gas fraction and a cored DM profile.  We first run the simulation until z = 6, when we insert five
black holes of mass $5\times10^5$ M$_{\odot}$ in the centers of the
five most massive halos at the time, which all have a mass of
$2\times10^8$ M$_{\odot}$ or higher. In these simulations black holes
do not accrete or produce feedback, as we are only
  interested in following their dynamics. From z$=$6 we run two
simulations to $z < 1$, one with our DF routine and the other
with advection. By the end of the simulation, there are only two major
star forming galaxies with masses $\sim10^{10}$ M$_{\odot}$.

We then run the Amiga Halo Finder (AHF) \citep{knollmann09} for all
  the saved snapshots and calculate the center of the main halo at each step using the shrinking spheres approach. In Figure 3 we
follow the trajectories of four black holes with respect to the center of this halo (which originally just has the blue BH at its center). Each color represents a different black hole. Black dots indicate a black hole
merger, which happens when two BHs come within two $\epsilon_{g}$ of one another at relative speeds low enough to be
gravitationally bound. The dashed black
lines indicate the gravitational softening length of 87 pc.

The black holes become perturbed as the red, cyan, and blue host galaxies
interact between $z = 3$ and $4$.  In the advection case the
black holes are driven quickly toward the center where they all merge,
leaving only one black hole (labeled as red).
With the DF correction, only the cyan and blue BHs
merge. After the red and blue galaxy hosts merge with DF, the blue and red BHs
remain orbiting around the center of the merger
remnant. The blue BH comes back to the center only after $4$ Gyr and
the red BH remains orbiting at around $1$ kpc ($11$ times the force resolution) for another $4$ Gyr before sinking and merging with the blue BH.

The more striking difference between the DF and the
  advection run involves the green black hole. When the much smaller green host
galaxy merges with the blue/red host at $z\sim1$, it is initially far from halo center
($\sim 30$ kpc) and is quickly disrupted by the main galaxy. With DF, the BH stays on a wide orbit, never
coming much closer than a few kpc from halo center. In the advection
case, however, the green black hole is quickly pushed to the center
where it merges with the central red BH (see Figures 3 and 4). This is an unrealistic
result, as the DF timescale of a $5\times10^5$ M$_{\odot}$ black hole that far
away from the center of such a small galaxy would be longer than a
Hubble time.

With this simulation we can clearly see how the choice of dynamical correction can affect the ability of SMBHs to become perturbed during mergers. In the DF simulation, BHs are able to remain off-center for many Gyr while with advection they are quickly driven to the center. Additionally, the DF simulation allows for sustained wide orbits resulting from minor mergers. Such dynamics can have an important impact on interpreting the connection between the initial occupation probability of SMBH seeds in dwarf galaxy progenitors and the observed occupation at low redshift. Methods such as the advection scheme presented here would predict a more direct connection, while the simulation with DF indicates that the nature of the mergers (i.e. mass ratio and orientation) can have an impact on which dwarf galaxies have observable SMBH activity and when that activity occurs. The DF correction could also have important implications for BH merger rates, allowing them to become more decoupled from galaxy merger rates than advection simulations. This will affect predictions of gravitational wave detections as well as estimates for recoiling BHs, which can have an important effect on observability \citep[e.g.][]{madau04}.

This simulation is a useful illustration of the variety of different, realistic BH orbits our method allows compared to commonly used advection schemes. In future work, we will explore the dynamics of BHs in a variety of different merger events within a cosmological context.

 \section{Discussion and Summary}
We have introduced a sub-grid force correction term for SMBH motion based on
   dynamical friction. This correction allows us to better model the orbital decay of
   SMBHs in numerical simulations.  We have shown using controlled
   experiments of isolated DM halos that this addition
   matches analytic predictions of the orbital decay in DM
   halos with resolutions attainable by large-volume cosmological
   simulations. We have also demonstrated that
   our prescription naturally converges with resolution.

   This method is a significant improvement over existing `advection'
   methods that force a short orbital decay timescale regardless of
   the dynamical state of the system. When applied to a
   cosmological dwarf galaxy simulation, our method results in
   noticeably different black hole dynamics compared with the
   advection scheme. In particular, our prescription:
\\
 
 $\bullet$ Models the perturbation and gradual orbital decay of a
 central BH during a galaxy merger
 \\
 
 $\bullet$ Allows for long-lived dual BH systems with close ($<1$ kpc) orbits.
 \\

 $\bullet$ Maintains a stable central BH when appropriate.
 \\
 
$\bullet$ Allows for sustained wide ($>5 $ kpc) orbit BHs.
 \\

 Correctly modeling the rich orbital dynamics of a black hole within
 its host galaxy can have important consequences for its accretion 
 history, duty cycle and observability that was
 previously neglected in simplified `advection' schemes. The
 dynamically complex and more realistic orbits allowed by our method will have
 crucial implications for the early growth of SMBHs, which takes place
 in small, rapidly growing galaxies at high redshift \citep{aykutalp14}. Additionally,
 understanding the relative importance of different accretion
 mechanisms throughout a SMBH's lifetime requires the ability to
 accurately model its dynamics during all phases of galaxy evolution,
 including merger events. Implementation of DF routines such as the
 one presented here will improve the ability of cosmological
 simulations to accurately model SMBH accretion, growth and energy deposition in the IGM and, therefore,
 increase the ability  of simulations to interpret and predict observational
 results. The implementation of this approach in our future cosmological volume and zoom simulations represents an exciting chance to realistically study the growth and merger rate of SMBHs across cosmic time. \\
\\
\\
FG and TQ were funded by NSF grant AST-0908499.  FG
acknowledges support from NSF grant AST-0607819 and NASA ATP
NNX08AG84G. MV acknowledges support from SAO award TM1-12007X, from NSF 
award AST 1107675, and from a Marie Curie FP7- Reintegration-Grant within the 7th European 
Community Framework Programme (PCIG10-GA-2011-303609).
Some results were obtained using the analysis software
pynbody \citep{pynbody}. ChaNGa was developed with support from
National Science Foundation ITR grant PHY-0205413 to the University of
Washington, and NSF ITR grant NSF-0205611 to the University of
Illinois. Simulations were run using NCSA Bluewaters and NASA Pleiades computers. 
Resources supporting this work were provided by NSF PRAC Award 1144357 and the 
NASA High-End Computing (HEC) Program through the NASA Advanced Supercomputing 
(NAS) Division at Ames Research Center. We thank  Andrew Pontzen, Aycin Aykutalp, John Wise, Jillian Bellovary, and Alyson Brooks for stimulating discussions and a careful reading of the manuscript.

\bibliography{bibref_mjt.bib}
\bibliographystyle{mn2e}

\bsp

\label{lastpage}
\end{document}